\documentclass[12pt]{article} 
\usepackage{graphicx}
\def \d {{\rm d}}

\def\boxit#1{\vbox{\hrule\hbox{\vrule\kern3pt
              \vbox{\kern3pt#1\kern3pt}\kern3pt\vrule}\hrule}}

\begin{document}

\title{\bf Accelerating and rotating black holes}

\author{J. B. Griffiths$^1$\thanks{E--mail: {\tt J.B.Griffiths@Lboro.ac.uk}} \ 
and J. Podolsk\'y$^2$\thanks{E--mail: {\tt Podolsky@mbox.troja.mff.cuni.cz}}
\\ \\ \small
$^1$Department of Mathematical Sciences, Loughborough University, \\
\small Loughborough,  Leics. LE11 3TU, U.K. \\
\small $^2$Institute of Theoretical Physics, Charles University in Prague,\\
\small V Hole\v{s}ovi\v{c}k\'ach 2, 18000 Prague 8, Czech Republic.}

\date{\today}
\maketitle

\begin{abstract}
\noindent
An exact solution of Einstein's equations which represents a pair of accelerating and rotating black holes (a generalised form of the spinning $C$-metric) is presented. The starting point is a form of the Pleba\'nski--Demia\'nski metric which, in addition to the usual parameters, explicitly includes parameters which describe the acceleration and angular velocity of the sources. This is transformed to a form which explicitly contains the known special cases for either rotating or accelerating black holes. Electromagnetic charges and a NUT parameter are included, the relation between the NUT parameter $l$ and the Pleba\'nski--Demia\'nski parameter $n$ is given, and the physical meaning of all parameters is clarified. The possibility of finding an accelerating NUT solution is also discussed. 
\end{abstract}


\section{Introduction}

The Schwarzschild and Kerr metrics, which describe the fields surrounding a static and rotating black hole respectively, are surely the best known solutions of Einstein's equations. Their charged versions are the Reissner--Nordstr\"om and Kerr--Newman solutions respectively. The $C$-metric, together with its charged version, is similarly well known. This represents a pair of causally separated black holes which accelerate away from each other under the action of forces represented by topological singularities along the axis. However, the inclusive solution, which represents a pair of (possibly charged) rotating and accelerating black holes has only recently begun to receive detailed attention. 

A large family of electrovacuum solutions of algebraic type~D was presented in 1976 by Pleba\'nski and Demia\'nski \cite{PleDem76}. This contains a number of parameters (including a cosmological constant). Special cases of this, after different transformations, include both the Kerr--Newman solution and the $C$-metric. In this sense, it must therefore include a solution which represents an accelerating and rotating black hole. When some of the Pleba\'nski--Demia\'nski parameters are set to zero, a solution is obtained that has come to be known as the ``spinning $C$-metric''. This has been studied by many authors (\cite{FarZim80b}--\cite{Pravdas02}). However, it has very recently been shown by Hong and Teo \cite{HongTeo05} that a different choice of parameters, which removes the properties associated with a non-zero NUT parameter, is more appropriate to represent an accelerating and rotating pair of black holes. 

In this paper, we present the family of solutions which describes the general case of a pair of accelerating and rotating charged black holes in a very convenient form. A generally non-zero NUT parameter is included, but the cosmological constant is taken to be zero. In appropriate limits, this family of solutions explicitly includes both the Kerr--Newman--NUT solution and the $C$-metric, without the need for further transformations. Also, the relation between the Pleba\'nski--Demia\'nski parameter $n$ and the NUT parameter $l$ is given explicitly, confirming that these should not generally be identified. This result enables us to discuss the possibility of accelerating NUT solutions, in addition to the accelerating and rotating solution with no NUT parameter that has been identified by Hong and Teo \cite{HongTeo05}. 

Our analysis starts with a new form of the Pleba\'nski--Demia\'nski metric which includes explicit parameters representing the acceleration of the sources and the twist of the repeated principal null congruences. These parameters are much more helpful than the traditional ones in determining the physical properties of this family of solutions.

\section{An initial form of the metric}

For the case in which the cosmological constant is zero, the Pleba\'nski--Demia\'nski metric \cite{PleDem76} (see also \S21.1.2 of \cite{SKMHH03}) is given by 
  \begin{equation} 
 \d s^2={1\over(1-\hat p\hat r)^2} \Bigg[
{{\cal Q}(\d\hat\tau-\hat p^2\d\hat\sigma)^2\over\hat r^2+\hat p^2} -{{\cal P}(\d\hat\tau+\hat r^2\d\hat\sigma)^2\over\hat r^2+\hat p^2} 
 -{\hat r^2+\hat p^2\over{\cal P}}\,\d\hat p^2
-{\hat r^2+\hat p^2\over{\cal Q}}\,\d\hat r^2 \Bigg] 
  \label{oldPDMetric}
  \end{equation}  
  where \ ${\cal P} =\hat k +2\hat n\hat p -\hat\epsilon\hat p^2 
  +2\hat m\hat p^3-(\hat k+\hat e^2+\hat g^2)\hat p^4$, \ 
  ${\cal Q} =\hat k+\hat e^2+\hat g^2 -2\hat m\hat r +\hat\epsilon\hat r^2 
  -2\hat n\hat r^3-\hat k\hat r^4$ \ 
  and $\hat m$, $\hat n$, $\hat e$, $\hat g$, $\hat\epsilon$ and $\hat k$ are arbitrary real parameters. It is usually assumed that $\hat m$ and $\hat n$ are the mass and NUT parameters although, as will be shown below, this is not generally the case. The parameters $\hat e$ and $\hat g$ represent electric and magnetic charges.

In the case when the parameter $\hat n$ and the charge parameters vanish (i.e. when $\hat n=\hat e=\hat g=0$) with $\hat k>0$ and $\hat\epsilon>0$, the metric (\ref{oldPDMetric}), or a transformation of it, is traditionally referred to as the ``spinning $C$-metric''. It is considered to represent two uniformly accelerating, rotating black holes, either connected by a conical singularity, or with conical singularities extending from each to infinity. Farhoosh and Zimmerman \cite{FarZim80b} have investigated the properties of the horizons for these solutions. Bi\v{c}\'ak and Pravda \cite{BicPra99} have also transformed the metric into Lewis--Papapetrou form and then expressed it in the canonical form of radiative space-times with boost-rotation symmetry. However, very recently, Hong and Teo \cite{HongTeo05} have shown that a different non-zero choice of $\hat n$ is more appropriate to describe a pair of accelerating and rotating black holes. (This will be analysed in greater detail below.)

With the metric in the form (\ref{oldPDMetric}), however, the acceleration and rotation of the sources are not clearly represented. For ease of interpretation, it is in fact convenient to introduce the rescaling 
  \begin{equation} 
  \hat p=\sqrt{\alpha\omega}\,p, \qquad \hat r=\sqrt{\alpha\over\omega}\,r, \qquad  \hat\sigma=\sqrt{\omega\over\alpha^3}\,\sigma, \qquad 
 \hat\tau=\sqrt{\omega\over\alpha}\,\tau,
  \label{scaling}
  \end{equation} 
 with the relabelling of parameters 
  \begin{equation} 
  \hat m+i\hat n=\Big({\alpha\over\omega}\Big)^{3/2}(m+in), \qquad
  \hat e+i\hat g={\alpha\over\omega}(e+ig), \qquad
 \hat\epsilon={\alpha\over\omega}\,\epsilon, \qquad
 \hat k=\alpha^2k. 
  \label{scaleps}
  \end{equation} 
 This introduces two additional parameters $\alpha$ and $\omega$ and the associated freedom to choose any two of the parameters in a more convenient way. With these changes, the metric becomes 
  \begin{equation}
  \begin{array}{r}
{\displaystyle  \d s^2={1\over(1-\alpha pr)^2} \Bigg[
{Q\over r^2+\omega^2p^2}(\d\tau-\omega p^2\d\sigma)^2
  -{P\over r^2+\omega^2p^2}(\omega\d\tau+r^2\d\sigma)^2 } \hskip3pc \\[12pt]
  {\displaystyle -{r^2+\omega^2p^2\over P}\,\d p^2
-{r^2+\omega^2p^2\over Q}\,\d r^2 \Bigg],}
  \end{array}
  \label{PleDemMetric}
  \end{equation}
  where
  \begin{equation}
  \begin{array}{l}
  P=P(p) =k +2\omega^{-1}np -\epsilon p^2 +2\alpha mp^3
-\alpha^2(\omega^2 k+e^2+g^2)p^4, \\[8pt]
  Q=Q(r) =(\omega^2k+e^2+g^2) -2mr +\epsilon r^2 -2\alpha\omega^{-1}nr^3
-\alpha^2kr^4,
  \end{array}
  \label{PQeqns}
  \end{equation}
  and $m$, $n$, $e$, $g$, $\epsilon$, $k$, $\alpha$ and $\omega$ are arbitrary real parameters. (It will be shown that the coefficient $\omega^{-1}n$ is in fact well behaved in the limit as $\omega\to0$.)

In terms of a convenient tetrad, the only non-zero component of the Weyl tensor is given by
  \begin{equation}
  \Psi_2= \left( -(m+in) +(e^2+g^2) {1+\alpha pr\over r-i\omega p} \right) 
\left({1-\alpha pr\over r+i\omega p}\right)^3.
  \label{Weyl1}
  \end{equation}
This confirms that these space-times are of algebraic type~D. The only non-zero component of the Ricci tensor is
  \begin{equation}
  \Phi_{11}= {1\over2}\,(e^2+g^2)\,{(1-\alpha pr)^4\over(r^2+\omega^2p^2)^2}.
  \label{Ricci1}
  \end{equation}
 These components indicate the presence of a curvature singularity at $r=0$, $p=0$. This singularity may be considered as the source of the gravitational field. They also show that the line element (\ref{PleDemMetric}) is flat if $m=n=0$ and $e=g=0$. (Notice that the remaining kinematical parameters $\epsilon$, $k$, $\alpha$ and $\omega$, may be non-zero in this flat limit.)

To retain a Lorentzian signature in (\ref{PleDemMetric}), it is necessary that $P>0$. Thus, the coordinate $p$ must be restricted to a particular range between appropriate roots of $P$. Specifically, if it is required that a singularity should appear in the boundary of the space-time, then this range must include $p=0$. This would require that $k>0$. However, important non-singular solutions also exist for which the chosen range of $p$ does not include $p=0$.

The points at which $P=0$ correspond to poles of the coordinates. By contrast, surfaces on which $Q=0$ are (Killing) horizons through which coordinates can be extended. Moreover, since \ $Q(r)=-\alpha^2r^4P(1/\alpha r)$, \ it is clear that $P$ and $Q$ have the same number of roots. In this paper, it will be assumed that these quartics have four distinct roots. However, other cases, which have less physical significance, also exist. Together with the requirement that $P>0$, the existence of four roots places certain restrictions on the ranges and signs of some of the parameters introduced as coefficients in (\ref{PQeqns}). In the case considered here, the parameters acquire the following physical interpretations:

[$e,g$]: By examining the electromagnetic field and the curvature tensor components, it is clear that $e$ and $g$ denote the electric and magnetic charges of the sources.

[$\alpha,\omega$]: Previously, it has been considered appropriate to use the rescaling (\ref{scaling}) with (\ref{scaleps}) to set $\alpha=1$ and $\omega=1$.  However, the physical interpretation of the solution can be determined more transparently by retaining $\alpha$ and $\omega$ as continuous parameters. In fact, $\alpha$ generally represents the acceleration of the sources. Also, $\omega$ is proportional to the twist of the repeated principal null directions and this relates to both the angular velocity of the sources and the NUT-like properties of the space-time. (It may be noted that these parameters can always be taken to be non-negative. If $\alpha<0$, its sign can always be changed by changing the signs of either $p$ and $n$ or $r$ and $m$. Similarly, if $\omega<0$, its sign can be changed by changing that of either $\sigma$ or $\tau$.) This approach also enables us to obtain the non-accelerating and non-rotating cases by trivially setting either of these parameters to zero.

[$\epsilon,k$]: Retaining $\alpha$ and $\omega$ as continuous parameters, we are free to use the rescaling (\ref{scaling}) with (\ref{scaleps}) in some other way. Actually, it is most convenient to use this freedom to rescale the parameters $\epsilon$ and $k$ to some specific values, although it is not possible to alter their signs. In fact $\epsilon$ and $k$ are related to the discrete parameters that specify the curvature of appropriate 2-surfaces and a canonical choice of coordinates on them. As appropriate for solutions representing black holes, we will only consider the case in which the relevant 2-surfaces have positive curvature. The signs of $\epsilon$ and $k$ have to be consistent with this interpretation.

[$m,n$]: For certain choices of the other parameters, it will be shown that $m$ is related to the mass of the source and $n$ to the NUT parameter. However, it should be emphasised that these only acquire their usual specific well-identified meanings in certain special sub-cases.

\section{A more general line element}

It can be shown that, with a transformation of the coordinate $\tau$, the line element (\ref{PleDemMetric}) already contains the Kerr--Newman solution when $\alpha=0$ and the $C$-metric when \hbox{$\omega=0$}. However, to introduce an explicit NUT parameter, and hence to include the non-singular NUT solution \cite{NewTamUnt63}, it is necessary to include a shift in the coordinate~$p$. To include this possibility, we start with the metric (\ref{PleDemMetric}) with (\ref{PQeqns}), and perform the coordinate transformation 
  \begin{equation}
  p=\omega^{-1}(l+a\tilde p), \qquad  \tau=t-(l+ap_0)^2a^{-1}\phi, 
  \qquad \sigma=-\omega a^{-1}\phi,
  \label{trans1A}
  \end{equation}
  where $a$, $l$ and $p_0$ are arbitrary parameters. By this procedure, we obtain the metric  
  \begin{equation}
  \begin{array}{l}
{\displaystyle \d s^2={1\over\Omega^2}\left\{
{Q\over\rho^2}\left[\d t- \left(a(p_0^2-\tilde p^2)
+2l(p_0-\tilde p) \right)\d\phi \right]^2
   -{\rho^2\over Q}\,\d r^2 \right.
} \\[8pt]
  \hskip8pc {\displaystyle 
 \left. -{\tilde P\over\rho^2} \Big[ a\d t  
  -\Big(r^2+(l+ap_0)^2\Big)\d\phi \Big]^2  
-{\rho^2\over\tilde P}\,\d\tilde p^2 \right\}, }
\end{array}
  \label{altPlebMetric}
  \end{equation}
  where
  $$ \begin{array}{l}
  \Omega=1-\alpha\omega^{-1} (l+a\tilde p)r \\[6pt]
  \rho^2 =r^2+(l+a\tilde p)^2 \\[6pt]
  \tilde P= a_0 +a_1\tilde p +a_2\tilde p^2
+ a_3\tilde p^3 +a_4\tilde p^4 \\[8pt]
  Q= (\omega^2k+e^2+g^2) -2mr +\epsilon r^2 
-2\alpha\omega^{-1}n r^3 -\alpha^2kr^4,
  \end{array} $$
  and we have put
  $$ \begin{array}{l}
  a_0= a^{-2} \Big( \omega^2k +2nl -\epsilon l^2
+2\alpha\omega^{-1} ml^3 -\alpha^2\omega^{-2}(\omega^2k+e^2+g^2)l^4 \Big) \\[6pt]
  a_1=2a^{-1}\Big( n -\epsilon l
+3\alpha\omega^{-1} ml^2 -2\alpha^2\omega^{-2}(\omega^2k+e^2+g^2)l^3 \Big) \\[6pt]
  a_2= -\epsilon +6\alpha\omega^{-1} ml -6\alpha^2\omega^{-2}(\omega^2k+e^2+g^2)l^2 \\[6pt]
  a_3= 2a\alpha\omega^{-1}m -4a\alpha^2\omega^{-2}(\omega^2k+e^2+g^2)l \\[6pt]
  a_4= -a^2\alpha^2\omega^{-2}(\omega^2k+e^2+g^2).
  \end{array} $$

We can now choose the new parameter $p_0$ to coincide with one of the roots of $\tilde P$. The metric (\ref{altPlebMetric}) is then regular at the pole $\tilde p=p_0$ which corresponds to an axis, so that it is appropriate to take $\phi$ as a periodic coordinate. We now consider the case in which $\tilde P(0)>0$ (i.e. $a_0>0$) for which a second root of $\tilde P$ will exist with the opposite sign to that of $p_0$. It is then possible to use the freedom to constrain the parameters $\epsilon$, $k$, $a$ and $l$ to fix this other root at $\tilde p=-p_0$. It can then be seen that the metric component $a(p_0^2-\tilde p^2)$ is regular at this second pole while the component $2l(p_0-\tilde p)$ is not. Thus, the metric is regular on the half-axis $\tilde p=p_0$, but a singularity of some kind occurs on the other half-axis $\tilde p=-p_0$. In fact, with this choice, $a$ corresponds to a Kerr-like rotation parameter for which the corresponding metric components are regular on the entire axis, while $l$ corresponds to a NUT parameter for which the corresponding components are only regular on the half-axis $\tilde p=p_0$.

In addition, the freedom in the choice of $\epsilon$, $k$, $a$ and $l$ can be further used to set $p_0=1$, so that $\tilde P$ has factors $(1-\tilde p)$ and $(1+\tilde p)$. With this, we can write 
 $$ \tilde P= (1-\tilde p^2)(a_0-a_3\tilde p-a_4\tilde p^2), $$ 
 so that the conditions  
 \begin{equation}
 a_1+a_3=0, \qquad a_0+a_2+a_4=0 
 \label{Pconditions}  
 \end{equation}
 must also be satisfied. The available freedom is sufficient to satisfy these two conditions provided the signs of $\epsilon$ and $k$ are consistent. In fact they provide two linear equations which specify the two parameters $\epsilon$ and $n$ as 
 \begin{eqnarray}
 &&\epsilon= {\omega^2k\over a^2-l^2}+4{\alpha l\over\omega}\,m 
 -{\alpha^2(a^2+3l^2)\over\omega^2}(\omega^2k+e^2+g^2), \label{epsilon}\\[8pt]
 &&n= {\omega^2k\,l\over a^2-l^2} -{\alpha(a^2-l^2)\over\omega}\,m 
 +{\alpha^2(a^2-l^2)l\over\omega^2}\,(\omega^2k+e^2+g^2). \label{n}  
 \end{eqnarray}
 Equation (\ref{n}) explicitly relates the Pleba\'nski--Demia\'nski parameter $n$ to the NUT parameter $l$, while (\ref{epsilon}) specifies the required value of $\epsilon$. With these, we obtain that 
 $$ a_0={\omega^2k\over a^2-l^2} -2{\alpha l\over\omega}\,m 
 +3{\alpha^2l^2\over\omega^2}(\omega^2k+e^2+g^2). $$ 
 In fact, it is also possible to choose the parameters such $a_0=1$, provided this is consistent with the signs of $\epsilon$ and $k$. (Otherwise this would imply that $\tilde P$ is not positive between roots at $\pm1$, the geometry would be different and the space-times would not represent black hole-like objects.) The required value of $k$ to achieve $a_0=1$ is given by 
  \begin{equation}
  \left( {\omega^2\over a^2-l^2}+3\alpha^2l^2 \right)\,k =1 +2{\alpha l\over\omega}\,m 
  -3{\alpha^2l^2\over\omega^2}(e^2+g^2). 
  \label{k}
  \end{equation}

The original metric (\ref{oldPDMetric}) contained two kinematical parameters $\hat\epsilon$ and $\hat k$. In the above argument, we have increased these parameters to $\epsilon$, $k$, $\alpha$, $\omega$, $a$ and $l$, but we have introduced three constraints that are effectively represented by (\ref{epsilon}), (\ref{n}) and (\ref{k}). One remaining freedom is therefore still available which could, for example, be used to set $\omega=|a+l|$.

\section{A new form of the metric}

With the above conditions satisfied, it is natural to put $\tilde p=\cos\theta$, where $\theta\in[0,\pi]$, so that $\theta$ spans the permitted range of $\tilde P$ between the roots $\tilde p=\pm1$. Substituting also for $\epsilon$ and $n$, the metric (\ref{altPlebMetric}) becomes 
  \begin{equation}
  \begin{array}{l}
{\displaystyle \d s^2={1\over\Omega^2}\left\{
{Q\over\rho^2}\left[\d t- \left(a\sin^2\theta
+4l\sin^2{\textstyle{\theta\over2}} \right)\d\phi \right]^2
   -{\rho^2\over Q}\,\d r^2 \right.
} \\[8pt]
  \hskip8pc {\displaystyle 
 \left. -{\tilde P\over\rho^2} \Big[ a\d t  
  -\Big(r^2+(a+l)^2\Big)\d\phi \Big]^2  
-{\rho^2\over\tilde P}\sin^2\theta\,\d\theta^2 \right\}, }
\end{array}
  \label{newMetric}
  \end{equation}
  where 
  \begin{equation}
  \begin{array}{l}
  {\displaystyle \Omega=1-{\alpha\over\omega}(l+a\cos\theta)\,r } \\[6pt]
  \rho^2 =r^2+(l+a\cos\theta)^2 \\[6pt]
  \tilde P= \sin^2\theta\,(1-a_3\cos\theta-a_4\cos^2\theta) \\[6pt]
  Q= {\displaystyle \left[(\omega^2k+e^2+g^2)\bigg(1+2{\alpha l\over\omega}\,r\bigg) 
  -2mr +{\omega^2k\over a^2-l^2}\,r^2\right] } \\[6pt]
   \hskip5pc \times {\displaystyle 
   \left[1+{\alpha(a-l)\over\omega}\,r\right] \left[1-{\alpha(a+l)\over\omega}\,r\right] }
  \end{array} 
  \label{newMetricFns}
  \end{equation} 
  and 
 \begin{equation}
 \begin{array}{l}
  {\displaystyle a_3= 2{\alpha a\over\omega}m -4{\alpha^2 al\over\omega^2}
  (\omega^2k+e^2+g^2) } \\[6pt]
  {\displaystyle a_4= -{\alpha^2a^2\over\omega^2}(\omega^2k+e^2+g^2) }
  \end{array}
 \label{a34}
 \end{equation} 
 with $k$ given by (\ref{k}). This contains seven arbitrary parameters $m$, $l$, $e$, $g$, $a$, $\alpha$ and $\omega$. Of these, the first six can be varied independently, and the remaining freedom can be used to set $\omega$ to any convenient value if at least one of the parameters $a$ or $l$ are non-zero.

The non-zero components of the curvature tensor are given by (\ref{Weyl1}) and (\ref{Ricci1}) in which $\omega p$ is replaced by $l+a\cos\theta$. It can be seen that, if $|l|\le|a|$, the metric (\ref{newMetric}) has a curvature singularity when $\rho^2=0$; i.e. at $r=0$, $\cos\theta=-l/a$. However, if $|l|>|a|$, it is non-singular.

 We present this metric as the family of solutions which represents a pair of accelerating and rotating charged black holes with a generally non-zero NUT parameter. We will now show that it reduces explicitly to familiar forms of either the Kerr--Newman--NUT solution or the $C$-metric in appropriate cases, without the need for further transformations. After that, we will consider further aspects of the interpretation of these solutions and some additional special cases.

\newpage

\section{Special cases}

It can first be seen that, when $\alpha=0$, we have \ $\omega^2k=a^2-l^2$ \ and hence 
 $$ \epsilon=1 \qquad \hbox{and} \qquad n=l. $$ 
 The second of these conditions simply identifies the Pleba\'nski--Demia\'nski parameter $n$ with the NUT parameter~$l$ in this case. Moreover, $\tilde P=\sin^2\theta$, and the metric (\ref{newMetric}) becomes 
  $$  \begin{array}{l}
{\displaystyle \d s^2= {Q\over\rho^2}\left[\d t- \left(a\sin^2\theta
+4l\sin^2{\textstyle{\theta\over2}} \right)\d\phi \right]^2
   -{\rho^2\over Q}\,\d r^2 } \\[8pt]
  \hskip8pc {\displaystyle 
 -{\sin^2\theta\over\rho^2} \Big[ a\d t  
  -\Big(r^2+(a+l)^2\Big)\d\phi \Big]^2 -\rho^2\,\d\theta^2,  }
\end{array} $$ 
  where
  $$ \begin{array}{l}
  \rho^2 =r^2+(l+a\cos\theta)^2 \\[6pt]
  Q= (a^2-l^2+e^2+g^2) -2mr + r^2 .
  \end{array} $$ 
 This is exactly the Kerr--Newman--NUT solution in the form which is regular on the half-axis $\theta=0$. It represents a black hole with mass~$m$, electric and magnetic charges $e$ and $g$, a rotation parameter $a$ and a NUT parameter~$l$.

Next, let us consider the case in which $\alpha$ is arbitrary but in which $l=0$ so that \ $\omega^2k=a^2$. \ In this case, it is convenient to use the remaining scaling freedom to put $\omega=a$, and hence 
 $$ \epsilon=1-\alpha^2(a^2+e^2+g^2), \qquad k=1, \qquad n=-\alpha am, \qquad \Omega=1-\alpha r\cos\theta. $$ 
 This will be discussed in detail in section~\ref{SpinC}. At this point, we will just consider the further limit in which $a\to0$. In this case, the metric (\ref{newMetric}) reduces to the form 
 $$ \d s^2={1\over(1-\alpha r\cos\theta)^2} \left( {Q\over r^2}\,\d t^2 -{r^2\over Q}\,\d r^2
  -\tilde P\,r^2\,\d\phi^2
 -{r^2\sin^2\theta\over\tilde P}\,\d\theta^2 \right), $$ 
 where 
  $$ \begin{array}{l}
  \tilde P=\sin^2\theta\Big(1-2\alpha m\cos\theta +\alpha^2(e^2+g^2)\cos^2\theta\Big), \\[6pt]
  Q=(e^2+g^2-2mr+r^2)(1-\alpha^2r^2).
  \end{array} $$ 
 This is exactly equivalent to the form for the charged $C$-metric that was introduced recently by Hong and Teo \cite{HongTeo03} using the coordinates $x=-\cos\theta$ and $y=-1/(\alpha r)$. It describes a pair of black holes of mass $m$ and electric and magnetic charges $e$ and $g$ which accelerate towards infinity under the action of forces represented by a topological (string-like) singularity, for which $\alpha$ is precisely the acceleration.

\section{Interpreting the new metric}

In view of the above limiting cases, it may be argued that (\ref{newMetric}) generally describes a pair of accelerating and rotating charged black holes, together with a NUT parameter. The specific relation between the Pleba\'nski--Demia\'nski parameter~$n$ and the NUT parameter~$l$ is determined by (\ref{n}) in which $k$ is given by (\ref{k}). In addition, the parameters $\alpha$ and $\omega$ are directly related to the acceleration and rotation of the sources respectively.

This interpretation certainly appears to be correct for the case where $|l|\le|a|$, in which the metric (\ref{newMetric}) has a curvature singularity at $r=0$, $\cos\theta=-l/a$. Since the space-time is asymptotically flat at conformal infinity, we take \ $r\in(0,r_\infty)$, \ where 
 $$ r_\infty= \left\{ 
 \begin{array}{cl}
 {\displaystyle {\omega\over\alpha(l+a\cos\theta)}} \qquad &\hbox{if} \quad a\cos\theta>-l \\[6pt]
 \infty \qquad &\hbox{otherwise}
 \end{array} \right. $$ 
 There is an acceleration horizon given by $\alpha r=\omega(|a|+l)^{-1}$. It may be noted that in this case the introduction of the parameter $l$ does not remove the curvature singularity as it does in the standard NUT solution.

By contrast, the case in which $|l|>|a|$ contains no curvature singularity. However, as the space-time is asymptotically flat at conformal infinity, we take \ $r\in(r_{-\infty},r_\infty)$, \ where 
 $$ \begin{array}{lcc}
\hbox{if} \ l>0, &r_{-\infty}=-\infty, \qquad
&{\displaystyle r_\infty={\omega\over\alpha(l+a\cos\theta)}} \\ 
\hbox{if} \ l<0, \qquad
&{\displaystyle r_{-\infty}={\omega\over\alpha(l+a\cos\theta)}},  \qquad
&r_\infty=\infty 
 \end{array} $$ 
 In this case, it can be seen from the form of $Q$ in (\ref{newMetricFns}) that outer and inner NUT-like horizons occur at $r=r_\pm$, where $r_\pm$ are the roots of the quadratic 
 $$ {\omega^2k\over a^2-l^2}\,r^2 
 -2\bigg(m-{\alpha l\over\omega}(\omega^2k+e^2+g^2)\bigg)r  
 +(\omega^2k+e^2+g^2) =0. $$ 
 The solution between the horizons ($r_-<r<r_+$) has a completely different structure to that of an accelerating Kerr black hole. It could represent a non-singular Taub-like vacuum cosmological model. There is also an acceleration horizon at $\alpha r=\omega(l\pm a)^{-1}$, taking whichever value is within the permitted range of~$r$.

We must now consider the regularity of the metric (\ref{newMetric}) on the axis for either of the above cases and for any value of $r$, although we will assume that we are working in the stationary regions.

Let us initially consider a small circle around the half-axis $\theta=0$ in the surface on which $t$ and $r$ are constant.  If we take $\phi\in[0,2\pi)$, we would obtain
 $$ {\hbox{circumference}\over\hbox{radius}} 
 =\lim_{\theta\to0} {2\pi\tilde P\over\rho^2} {(r^2+(a+l)^2)\over\theta\sin\theta} =2\pi(1-a_3-a_4), $$ 
 where $a_3$ and $a_4$ are given by (\ref{a34}). This would correspond to a conical singularity with a deficit angle of 
 $$ \delta=2\pi(a_3+a_4) ={2\pi\alpha a\over\omega} \left( 2m-{\alpha(4l+a)\over\omega}(\omega^2k+e^2+g^2) \right). $$ 
  When we come to consider the half-axis $\theta=\pi$, however, it must be noticed that, near the axis in the stationary regions in which $Q>0$, the lines on which $t$, $r$ and $\theta$ are constant are closed timelike lines if $l\ne0$.  We must therefore consider small circles on the surfaces on which $t'$ and $r$ are constant, where $t'=t-4l\phi$. In this case 
 $$ {\hbox{circumference}\over\hbox{radius}} 
 =\lim_{\theta\to\pi} {2\pi\tilde P\over\rho^2} 
 {(r^2+(a-l)^2)\over(\pi-\theta)\sin\theta} =2\pi(1+a_3-a_4). $$ 
  i.e. there is an excess angle in this case if $\phi\in[0,2\pi)$ and $a_3-a_4>0$.

Of course, as in any NUT solution, there are two distinct possible interpretations of the space-time. One, due to Misner \cite{Misner63}, is to consider $t$ as a periodic coordinate with period $8\pi l$, so that the space-time has topology $R\times S^3$. This may be appropriate in the time-dependent region, in which case the solution would represent a generalised Taub universe. The alternative approach, due to Bonnor \cite{Bonnor69}, is to treat the singularity at $\theta=\pi$ as a semi-infinite line singularity which acts as a spike injecting angular momentum into the source. In this case, the topological singularity in the stationary regions is surrounded by a region which contains closed timelike lines. (Bonnor \cite{Bonnor01} has described these as ``torsion singularities''.) Since the emphasis here is on the exterior stationary regions, we will generally adopt the second interpretation below.

By considering the $C$-metric limit, it can be seen that the half-axis $\theta=0$ is that which connects the two distinct black holes to infinity, while the half-axis $\theta=\pi$ corresponds to that between the two black holes. By adjusting the range of $\phi$, it is always possible to remove the conical singularity on either of these half-axes. For example, the singularity on $\theta=\pi$ can be removed by taking $\phi\in\big[0,2\pi(1+a_3-a_4)^{-1}\big)$. The acceleration of the two ``sources'' would then be achieved by two ``strings'' of deficit angle 
  \begin{equation}
 \delta_0 = {4\pi\,a_3 \over1+a_3-a_4} 
  \label{def0}
  \end{equation} 
 connecting them to infinity. Alternatively, the singularity on $\theta=0$ could be removed by taking $\phi\in\big[0,2\pi(1-a_3-a_4)^{-1}\big)$, and the acceleration would then be achieved by a ``strut'' between them in which the excess angle is given by 
  \begin{equation}
 -\delta_\pi = {4\pi\,a_3 \over1-a_3-a_4}. 
  \label{defpi}
  \end{equation}

It should also be recalled that, when $l\ne0$, the metric (\ref{newMetric}) has an additional singularity on $\theta=\pi$ which, in the general case, corresponds to the ``axis'' between the two ``sources''. However, this can be transformed to the other axis by the transformation $t'=t-4l\phi$. It can thus be seen that the topological singularity on the axis which causes the acceleration, and the singularity on the axis associated with the NUT parameter and the existence of closed timelike lines, are mathematically independent. They may each be set on whatever parts of the axis may be considered to be most physically significant.

\section{A pair of accelerating and rotating black holes}
\label{SpinC}

Let us now return to the case in which $l=0$ so that \ $\omega^2k=a^2$ \ and we may put $\omega=a$. As shown above, the conditions (\ref{epsilon}) and (\ref{n}) then imply that $\epsilon=1-\alpha^2(a^2+e^2+g^2)$ and $n=-\alpha am$. It can be seen explicitly that the Pleba\'nski--Demia\'nski parameter $n$ is non-zero in this case, while the NUT parameter $l$ vanishes. The resulting solution corresponds precisely to that of Hong and Teo \cite{HongTeo05} which represents an accelerating and rotating pair of black holes without any NUT-like behaviour.

In this case, the metric may be taken in the form 
  \begin{equation}
 \d s^2={1\over\Omega^2}\left\{
{Q\over\rho^2}\left[\d t- a\sin^2\theta\,\d\phi \right]^2
   -{\rho^2\over Q}\,\d r^2  -{\tilde P\over\rho^2} \Big[ a\d t  
  -(r^2+a^2)\d\phi \Big]^2  
-{\rho^2\over\tilde P}\sin^2\theta\,\d\theta^2 \right\}, 
  \label{HongTeoMetric}
  \end{equation}
  where
  $$ \begin{array}{l}
  \Omega=1-\alpha r\cos\theta \\[6pt]
  \rho^2 =r^2+a^2\cos^2\theta \\[6pt]
  \tilde P= \sin^2\theta \Big(1-2\alpha m\cos\theta +\alpha^2(a^2+e^2+g^2)\cos^2\theta\Big) \\[6pt]
  Q= \Big((a^2+e^2+g^2) -2mr +r^2\Big) (1-\alpha^2r^2).
  \end{array} $$ 
  The only non-zero components of the curvature tensor are given by
 $$  \begin{array}{l}
 {\displaystyle \Psi_2= \left(-m(1-i\alpha a)
+(e^2+g^2) {1+\alpha r\cos\theta\over r-ia\cos\theta} \right)
 \left({1-\alpha r\cos\theta\over r+ia\cos\theta}\right)^3 } \\[12pt]
 {\displaystyle \Phi_{11}= {1\over2}\,(e^2+g^2)\,{(1-\alpha r\cos\theta)^4\over(r^2+a^2\cos^2\theta)^2}.}
  \end{array} $$ 
  These indicate the presence of a Kerr-like ring singularity at $r=0$, $\theta={\pi\over2}$. Thus, we may generally take $r\in(0,r_\infty)$.

If $m^2\ge a^2+e^2+g^2$, the expression for $Q$ factorises as 
  $$  Q = (r_--r)(r_+-r)(1-\alpha^2r^2), $$ 
  where
  \begin{equation}
  r_\pm = m\pm\sqrt{m^2-a^2-e^2-g^2}.
  \label{KerrNewman roots}
  \end{equation}
 The expressions for $r_\pm$ are identical to those for the locations of the outer and inner horizons of the non-accelerating Kerr--Newman black hole. However, in this case, there is another horizon at $r=\alpha^{-1}$ which is already familiar in the context of the $C$-metric as an acceleration horizon.

However, the most significant advantage of the form (\ref{HongTeoMetric}) is that the range of $\theta$ is bounded by roots which correspond to $\tilde p=1$ and $\tilde p=-1$, and this ensures that closed timelike lines do not appear near the conical singularities on the axis -- a property that is usually associated with a non-zero value of the NUT parameter. This confirms that the correct ``spinning $C$-metric'' without any NUT-like behaviour occurs when $l=0$ (i.e. $n=-\alpha am$) as argued by Hong and Teo \cite{HongTeo05}.

In addition, we also obtain in this case that 
 $$ a_3=2\alpha m, \qquad a_4=-\alpha^2(a^2+e^2+g^2), $$ 
 so that the deficit angle of the string pulling the black holes toward infinity, or the excess angle of the strut between them, is obtained immediately using (\ref{def0}) or (\ref{defpi}) respectively.

\section{Apparently accelerating NUT solutions}

Let us now consider the complementary case with no Kerr-like rotation in which $a\to0$, but $l$ remains non-zero. In this case $a_3$ and $a_4$ vanish, and we can use the scalings (\ref{scaling}) to set the values of $\epsilon$ and $k$ such that $a_0=1$ and $a_2=-1$. The remaining freedom can then be used to set $\omega=l$, and (\ref{epsilon}) and (\ref{n}) become 
 $$ \begin{array}{l}
 \epsilon=-k+4\alpha m -3\alpha^2(e^2+g^2+kl^2) \\[6pt] 
 n= l\,\Big(-k+\alpha m -\alpha^2(e^2+g^2+kl^2)\Big), 
 \end{array} $$ 
 where
  $$ (1-3\alpha^2l^2)k=-1-2\alpha m+3\alpha^2(e^2+g^2). $$ 
 For the general case in which $\alpha\ne0$, this yields a fairly complicated relation between the parameter $n$ and the NUT parameter~$l$.

\goodbreak
We now have $\tilde P= \sin^2\theta$, and the metric (\ref{newMetric}) becomes 
  \begin{equation}
   \d s^2= \tilde Q \left(\d t -4l\sin^2{\textstyle{\theta\over2}}\,\d\phi \right)^2 
 -{\d r^2 \over\tilde Q\,(1-\alpha r)^4}
 -{r^2+l^2\over(1-\alpha r)^2} \left(\d\theta^2 +\sin^2\theta\,\d\phi^2 \right),
  \label{a=0Metric}
  \end{equation}
  where
  $$ \tilde Q= {(e^2+g^2+kl^2)(1+2\alpha r)-2mr-kr^2 \over r^2+l^2}. $$ 
  The only non-zero components of the curvature tensor are given by
 $$  \begin{array}{l}
 {\displaystyle \Psi_2= \left(-(m+in)
+(e^2+g^2) {1+\alpha r\over r-il} \right)
 \left({1-\alpha r\over r+il}\right)^3 } \\[12pt]
 {\displaystyle \Phi_{11}= {1\over2}\,(e^2+g^2)\,{(1-\alpha r)^4\over(r^2+l^2)^2}, }
  \end{array} $$ 
 from which it can be seen that there are no curvature singularities provided $l\ne0$, and we may generally take $r\in(r_{-\infty},r_\infty)$ as defined above.

The metric (\ref{a=0Metric}) appears to contain the arbitrary parameter $\alpha$. For the case in which $\alpha=0$, we obtain $k=-1$ and \ 
$\tilde Q=\big(r^2-2mr-l^2+e^2+g^2\big)/(r^2+l^2)$. \ This is precisely the (charged) NUT metric \cite{NewTamUnt63}. It is therefore natural to initially interpret the metric (\ref{a=0Metric}) as a general accelerating NUT solution. However, this is not the correct interpretation for this particular case.

It may first be observed that the deficit angles vanish on the entire axis \ $\theta=0,\pi$ \ in this case in which $a=0$; i.e. there is nothing to ``cause'' an acceleration. More specifically, the transformation 
 \begin{equation} 
 r={1\over\sqrt{1-\alpha^2L^2}}\left( {\sqrt{1-\alpha^2L^2}\,R-\alpha L^2\over\alpha R+\sqrt{1-\alpha^2L^2}} \right), \qquad 
 t={T\over\sqrt{1-\alpha^2L^2}}, \qquad l={L\over\sqrt{1-\alpha^2L^2}}, 
 \label{a=0Trans} 
 \end{equation}  
 alters the metric (\ref{a=0Metric}) to the form 
 $$  \d s^2= F \left(\d T -4L\sin^2{\textstyle{\theta\over2}}\,\d\phi \right)^2 
 -{\d R^2 \over F} 
 -(R^2+L^2) \left(\d\theta^2 +\sin^2\theta\,\d\phi^2 \right), $$ 
 where 
 $$ F={R^2-2MR-L^2+e^2+g^2\over R^2+L^2}, $$ 
 and 
 \begin{equation} 
 M= { m -2\alpha(1-\alpha^2L^2)(e^2+g^2) +(3-4\alpha^2L^2)\alpha L^2
  \over\sqrt{1-\alpha^2L^2}\>(1-4\alpha^2L^2)}. 
 \label{Mm}
 \end{equation}  
 This is exactly the charged NUT solution (when $L=0$, it is the usual form of the Reissner--Nordstr\"om metric). But the familiar mass parameter is now $M$, and this is related by equation (\ref{Mm}) to the Pleba\'nski--Demia\'nski parameters $m$, $e$ and $g$ and the parameters $L$ and $\alpha$. In addition, the NUT parameter is now given by $L$, and this is related to $\alpha$ and $l$ as in (\ref{a=0Trans}) and hence to the Pleba\'nski--Demia\'nski parameter $n$ which is given by (\ref{n}). It may also be observed that, in the above transformation, 
 $$ R={1\over\sqrt{1+\alpha^2l^2}}\left({r+\alpha l^2\over1-\alpha r}\right), $$ 
 so that $r=\alpha^{-1}$ corresponds to spacelike infinity in the line element (\ref{a=0Metric}). But, particularly, it must be concluded that $\alpha$ is a redundant parameter in the metric (\ref{a=0Metric}) that can be removed by the above transformation for this case in which $a=0$. Thus, the metric (\ref{a=0Metric}) does not represent a new ``accelerating'' NUT solution.

On the other hand, the general metric (\ref{newMetric}) does represent a pair of accelerating sources with a non-zero NUT parameter since it does explicitly contain the $C$-metric. And, just as the accelerating and rotating solution with no NUT features is obtained by putting \hbox{$n=-\alpha\omega m$} and not $n=0$, so it may be appropriate to look for an accelerating NUT solution with no rotational features by considering some non-zero value of $a$ rather than putting $a=0$. However, such a solution has to date neither been identified nor proved not to exist.

\section{Conclusions} 

We have presented, in a most convenient form, the metric (\ref{newMetric}) which explicitly represents the complete family of accelerating and rotating black holes with a generally non-zero mass, charge (electric and magnetic) and NUT parameter. The general structure of this family of solutions is indicated in figure~1.

\begin{figure}[hpt]
\begin{center} \includegraphics[scale=0.85, trim=5 5 5 -5]{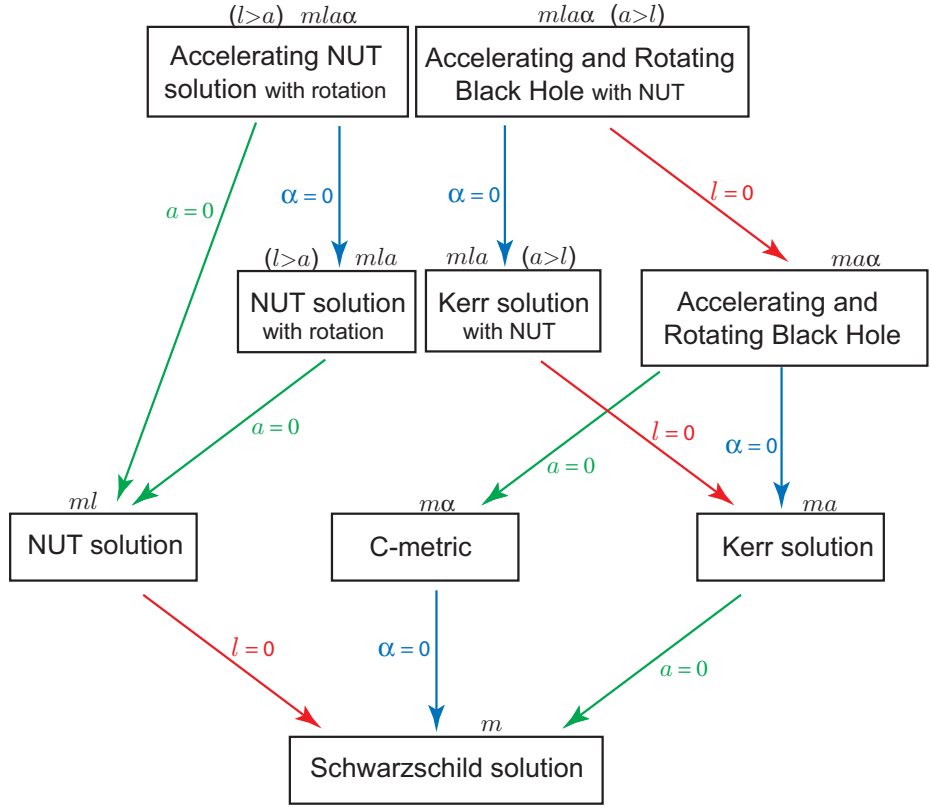} 
\caption{ \small The structure of the family of solutions presented here when $e=g=0$ and \hbox{$m\ne0$}. This family has four parameters $m$, $l$, $a$ and $\alpha$. We have distinguished an (accelerating) Kerr solution with a small NUT parameter from an (accelerating) NUT solution with a small rotation as their singularity structures differ significantly even though their metric forms (\ref{newMetric}) are identical. An accelerating NUT solution without rotation has not been identified. All special cases have obvious charged versions. }
\end{center}
\end{figure}

The metric has been obtained after first re-expressing the Pleba\'nski--Demia\'nski metric in a more convenient form which explicitly includes the parameters $\alpha$ and $\omega$, which respectively represent the acceleration of the sources and the twist of the repeated principal null congruences (and hence the rotation of the sources and/or the effects of the NUT parameter). The form (\ref{newMetric}) excludes the cases of the Pleba\'nski--Demia\'nski solution for which the spacelike 2-surfaces spanned by $p$ and $\sigma$ have zero or negative curvature, as these do not represent black hole-like solutions in any limit.

In presenting this form of the metric, we have clarified the physical meaning of the arbitrary parameters involved. As can be seen from the Maxwell field components, $e$ and $g$ are unambiguously the electric and magnetic charges of the sources. The twist of the repeated principal null congruences is proportional to the parameter $\omega$, and this is related to both the Kerr-like rotation parameter $a$ and the NUT parameter $l$. Provided $|a|>|l|$, $\alpha$ is the acceleration of the sources. In particular, we have found an explicit relationship between the parameter $l$ and the Pleba\'nski--Demia\'nski parameter $n$ given in (\ref{n}). However, the NUT parameter is only identified as $l$ for the non-accelerating case in which $\alpha=0$. When $\alpha\ne0$, it may be more appropriately taken as $L$ which is related to $l$ by (\ref{a=0Trans}). But even this is only valid when $a=0$. Thus the all-encompassing physical representation of the NUT parameter when $\alpha$ and $a$ are both non-zero, is still to be determined.

It is also difficult to physically determine the ``correct'' mass parameter. When $n=0$, the gravitational strength of the singularity is proportional to $m$, but otherwise it is proportional to $\sqrt{m^2+n^2}$. However, when $|l|>|a|$, the space-time is nonsingular and the ``mass'' of the source is only determined in certain limits. When $\alpha=0$, $m$ is still the familiar mass parameter of the Kerr--Newman--NUT metric. However, when $a=0$, the usual mass parameter of the charged $C$-metric is $M$, which given by (\ref{Mm}). Thus, as for the NUT parameter, the physically significant mass parameter is still to be determined when $\alpha$, $L$ and $a$ are all non-zero.

\section*{Acknowledgements}

This work was supported in part by a grant from the EPSRC.

\end{document}